# Theoretical Analysis of Functionally Graded Piezoelectric Thick-walled Cylinder Subjected to Mechanical and Electric Loadings

Han Wang [a], Libiao Xin [a, *], Guansuo Dui [b.]

[a] Institute of Applied Mechanics, College of Mechanical and Vehicle Engineering, Taiyuan University of Technology, Taiyuan, 030024, China

[b] Institute of Mechanics, Beijing Jiaotong University, Beijing, 100044, China

## Abstract

In this paper, the theoretical analysis for a hollow thick-walled functionally graded piezoelectric cylinder subjected to electric and mechanical loads are developed. The cylinder consists of two materials (PZT4 and PVDF) and the volume fraction of PZT4 is given in the three variable parameters power law form. By using the Voigt method and the assumption of a uniform strain field within the two linear elastic constituents, the complex hypergeometric differential equation of the radial displacement is obtained. Then the solutions of the radial displacement, the stresses, and the electric potential are derived and solved. The method in this paper is more suitable for actual engineering gradient piezoelectric materials, and the volume fraction function can cover more complicated situations. Finally, the influence of the parameter *n* in volume fraction on the mechanical behaviors are investigated, and the difference between the circumferential and radial stresses is discussed to reduce the stress concentration in the functionally graded piezoelectric cylinder.

---

* Corresponding author, E-mail: xinlibiao@tyut.edu.cn

**Nomenclature**

| | |
|---|---|
| $a$, $b$ | inner and outer radii |
| $r$ | radial coordinate |
| $c(r)$ | volume fraction of material A |
| $c_0$, $k$, $n$ | material parameters in the volume fraction $c(r)$ |
| $p_a$, $p_b$ | internal and external pressures |
| $\varphi_a$, $\varphi_b$ | internal and external electric potentials |
| $C_{ijkl}^{i}$ ($i$=1, 2) | elastic modulus of the component |
| $e_{mij}^{i}$ ($i$=1, 2) | piezoelectric tensor modulus of the component |
| $k_{mk}^{i}$ ($i$=1, 2) | dielectric modulus of the component |
| $u$ | radial displacement |
| $\varphi$ | electric potential |
| $\varepsilon_r^{(i)}$, $\varepsilon_\theta^{(i)}$ | radial and circumferential strains of the component |
| $\varepsilon_r$, $\varepsilon_\theta$ | average radial and circumferential strains of the cylinder |
| $\sigma_r^{(i)}$, $\sigma_\theta^{(i)}$, $\sigma_z^{(i)}$ | radial, circumferential, and axial stresses of the component |
| $\sigma_r$, $\sigma_\theta$, $\sigma_z$ | average radial, circumferential, and axial stresses of the cylinder |
| $E_r^{(i)}$, $D_r^{(i)}$ | radial electric field and electric displacement of the component |
| $E_r$, $D_r$ | average radial electric field and electric displacement of the cylinder |
| $a(r)$, $b(r)$, $d(r)$, $g(r)$, $f(r)$, $h(r)$, $k(r)$ | functions related to the volume fraction $c(r)$ |
| $x$ | new variables about $r$ |
| $F$ | Hypergeometric function |
| $\alpha, \beta, \delta$ | constants in hypergeometric function |
| $\phi_i$ ($i$=1,2,....,6) | coefficients in Eq. (19) |
| $P(r)$, $Q(r)$ | two linearly independent solutions of Eq. (22) |
| $G(r)$ | a particular solution of Eq. (22) |
| $\bar{\alpha}$, $\bar{\beta}$, $\bar{\delta}$ | constants in $G(r)$ |
| $A(r)$, $B(r)$, $C(r)$ | functions in electric potential |
| $C_i$ ($i$=1,2,....,4) | integral constants |
| $C_{11}$, $C_{12}$, $C_{21}$, $C_{22}$, $C_0$, $\bar{C}_0$ | constants in integral constants |
| $\bar{r}$ | non-dimensional radial coordinate |
| $\bar{a}$, $\bar{b}$ | non-dimensional inner and outer radii |
| $\bar{u}$ | non-dimensional radial displacement |
| $\bar{\sigma}_{ij}$ | non-dimensional stresses |
| $\bar{\varphi}$ | non-dimensional electric potential |

## 1. Introduction

Piezoelectric materials are smart materials that can generate electrical activity in response to minute deformation, which is widely used due to their direct and inverse effect in various electronic applications such as transducers, sensors, actuators, and biological equipment recently [1-3]. However, these applications are greatly limited in some respects due to the shortcomings of piezoelectric materials such as its brittleness and low power consumption. In the last several decades, functionally graded materials (FGMs) have been thoroughly investigated because of their ability to optimize mechanical behaviors by setting the material parameters as some unique functional forms. FGMs have been applied in many engineering fields: energy conversion systems, transport systems, cutting tools, machine parts, semiconductors, surface wrinkling in FGM elastomer and biosystems [4-6]. Therefore, with such attractive and practical advantages, the combination of the FGMs and the piezoelectric materials produces functionally graded piezoelectric materials (FGPMs) with more comprehensive applications.

As for the functionally graded piezoelectric cylinders under various loading conditions, there have been a large number of researchers exploring its electroelastic responses by using different methods. Authors [7] obtained an exact solution for a thick-walled cylinder made of FGPM under thermal and mechanical loads by assuming power functions for all mechanical and thermal properties. For example, the elastic stiffness was assumed as $C_{ij}(r)=C_{ij}^{0}r^{l}$ , piezoelectric coefficient was assumed as $e_{ij}(r)=e_{ij}^{0}r^{l}$, and the dielectric constant was assumed as $\eta_{ij}(r)=\eta_{ij}^{0}r^{l}$ ($C_{ij}^{0}$, $e_{ij}^{0}$, $\eta_{ij}^{0}$ and $l$ are material constants, $r$ is the radial coordinate of the cylinder). The elastic behaviors of a hollow cylinder composed of functionally graded piezoelectric material, placed in a uniform magnetic field, subjected to electric, thermal and mechanical loads are presented by assuming the material parameters as power functions [8]. In the following years, similar power function assumptions can be found in many articles [9-18]. Some authors proposed different assumptions that [19-21] the material parameters were assumed as exponential function of the radius. For example, the elastic stiffness was $C_{ij}(r) = C_{ij}^{0} e^{\alpha}$, piezoelectric coefficient was $e_{ij}(r) = e_{ij}^{0} e^{\alpha}$, and the dielectric constant was $\eta_{ij}(r) = \eta_{ij}^{0} e^{\alpha}$ ($C_{ij0}$, $e_{ij0}$, $\eta_{ij0}$ and $\alpha$ are material constants).

In the above papers, material parameters (the elastic stiffness, piezoelectric coefficient and the dielectric constant) of FGPM are assumed as power functions or exponential functions, which means FGPM consists of multiple layers. This point may violate the concept of FGM formed of two or more constituent phases with a continuously variable composition. On the other hand, all material parameters have the same index value ( $l$ or $\alpha$), which seems unreasonable in practice. Furthermore, there are two variable constants in power or exponential functions, which are not sufficient for describing more complex cases.

In order to solve the above mentioned problems, from the concept of functionally graded materials at the beginning we consider that the cylinder consists of two-phase materials, giving the volume fraction of one phase of the material rather than the assumption of some material parameters of the FGPM cylinder. In this work, the volume fraction of one phase change with three variable parameters different from two variable parameters mentioned above. In most actual systems, the overall material parameters cannot be found directly, but they can be obtained in terms of their properties of constituents and the volume fractions in a certain regulation [22, 23]. The constitutive model is then obtained based on the Voigt method, and the electroelastic behaviors of the functionally graded piezoelectric cylinder are given. Finally, the influence of the parameter in volume fraction on the radial displacement, the stresses and the electric potential of the FGPM cylinder are studied.

**2. Theoretical analysis**

A hollow radially polarized functionally graded piezoelectric cylinder is considered here. Denote the thickness of this cylinder as *b*-*a* with inner radius *a* and outer radius *b*. It is assumed that this cylinder is sufficiently long and subjected to axisymmetric mechanical and electric loadings on its surfaces (Fig.1). Cylindrical polar coordinates $(r,\theta,z)$ are used and the stress boundary conditions are $\sigma_r|_{r=a} = -p_a$ and $\sigma_r|_{r=b} = -p_b$, the electric boundary conditions are $\varphi|_{r=a} = \varphi_a$ and $\varphi|_{r=b} = \varphi_b$.

The cylinder consists of two materials A (PZT4) and B (PVDF), however, the interaction between materials A and B is not considered in this work. The volume fraction $c(r) \in [0,1]$ of material A is given by

$$c(r) = c_0\left[1 - k(r/b)^n\right] \tag{1}$$

where *r* is the radius and $c_0$, *k* and *n* are the material parameters. We can achieve a wide range of nonlinear and continuous profiles to describe the reasonable evolution of the volume fraction *c*(*r*) in the functionally graded piezoelectric cylinder by choosing suitable *n* and *k* values. In particular, $c_0$=0 and $c_0$=1, *k*=0 denote the cylinder totally consists material B and material A, respectively.

As is known to all, material properties will show transversely isotropic properties if the polarization direction of the piezoelectric material is along a certain direction. For cylindrically orthotropic homogeneous piezoelectric materials polarized in the radial direction, the tensor forms of constitutive equations take the following form

$$\begin{aligned} \sigma_{ij}^{(i)} &= C_{ijkl}^{i}\varepsilon_{kl}^{(i)} - e_{mij}^{i}E_{m}^{(i)} \\ D_{m}^{(i)} &= e_{mij}^{i}\varepsilon_{ij}^{(i)} + k_{mk}^{i}E_{m}^{(i)} \end{aligned} \tag{2}$$

the subscript $i$=1, 2 denotes material A and B, respectively; $\sigma_{ij}^{(i)}$ and $\varepsilon_{kl}^{(i)}$, $\varepsilon_{ij}^{(i)}$ are the stress and strain tensors, respectively; $E_{m}^{(i)}$ and $D_{m}^{(i)}$ are electric field and electric displacement vectors, respectively; $C_{ijkl}^{i}$, $e_{mij}^{i}$ and $k_{mk}^{i}$ are elastic, piezoelectric tensor and dielectric moduli, respectively.

For convenience, the component forms of Eq. (2) for this cylinder can be written as

$$\begin{aligned} \sigma_{r}^{(i)} &= C_{13}^{i}\varepsilon_{\theta}^{(i)} + C_{33}^{i}\varepsilon_{r}^{(i)} - e_{33}^{i}E_{r}^{(i)} \\ \sigma_{\theta}^{(i)} &= C_{11}^{i}\varepsilon_{\theta}^{(i)} + C_{13}^{i}\varepsilon_{r}^{(i)} - e_{31}^{i}E_{r}^{(i)} \\ \sigma_{z}^{(i)} &= C_{12}^{i}\varepsilon_{\theta}^{(i)} + C_{13}^{i}\varepsilon_{r}^{(i)} - e_{31}^{i}E_{r}^{(i)} \\ D_{r}^{(i)} &= e_{31}^{i}\varepsilon_{\theta}^{(i)} + e_{33}^{i}\varepsilon_{r}^{(i)} + k_{33}^{i}E_{r}^{(i)} \end{aligned} \tag{3}$$

where $\sigma_{r}^{(i)}$, $\sigma_{\theta}^{(i)}$ and $\sigma_{z}^{(i)}$ are the radial, circumferential and axial stresses, respectively; $\varepsilon_{r}^{(i)}$ and $\varepsilon_{\theta}^{(i)}$ are the radial and circumferential strains, respectively; $E_{r}^{(i)}$ and $D_{r}^{(i)}$ are electric field and electric displacement in radial direction, respectively.

Eq. $(3)_4$ can be easily rearranged as

$$E_{r}^{(i)} = \frac{D_{r}^{(i)} - e_{31}^{i}\varepsilon_{\theta}^{(i)} - e_{33}^{i}\varepsilon_{r}^{(i)}}{k_{33}^{i}} \tag{4}$$

Substituting Eq. (4) into the stress components of Eq. (3), we have

$$
\begin{aligned}
\sigma_r^{(i)} &= \left(C_{13}^i + \frac{e_{31}^i e_{33}^i}{k_{33}^i}\right)\varepsilon_\theta^{(i)} + \left(C_{33}^i + \frac{e_{33}^{i\ 2}}{k_{33}^i}\right)\varepsilon_r^{(i)} - \frac{e_{33}^i}{k_{33}^i} D_r^{(i)} \\
\sigma_\theta^{(i)} &= \left(C_{11}^i + \frac{e_{31}^{i\ 2}}{k_{33}^i}\right)\varepsilon_\theta^{(i)} + \left(C_{13}^i + \frac{e_{31}^i e_{33}^i}{k_{33}^i}\right)\varepsilon_r^{(i)} - \frac{e_{31}^i}{k_{33}^i} D_r^{(i)} \\
\sigma_z^{(i)} &= \left(C_{12}^i + \frac{e_{31}^{i\ 2}}{k_{33}^i}\right)\varepsilon_\theta^{(i)} + \left(C_{13}^i + \frac{e_{33}^i e_{31}^i}{k_{33}^i}\right)\varepsilon_r^{(i)} - \frac{e_{31}^i}{k_{33}^i} D_r^{(i)}
\end{aligned} \tag{5}
$$

The average stress tensor, strain tensor, electric field tensor and electric displacement tensor over a representative volume element $V$ are defined as [24]

$$
\begin{aligned}
\boldsymbol{\sigma} &= \frac{1}{V}\int_V \hat{\boldsymbol{\sigma}}(x)dx, & \boldsymbol{\sigma}^{(i)} &= \frac{1}{V_i}\int_{V_i} \hat{\boldsymbol{\sigma}}^{(i)}(x)dx \\
\boldsymbol{\varepsilon} &= \frac{1}{V}\int_V \hat{\boldsymbol{\varepsilon}}(x)dx, & \boldsymbol{\varepsilon}^{(i)} &= \frac{1}{V_i}\int_{V_i} \hat{\boldsymbol{\varepsilon}}^{(i)}(x)dx \\
\boldsymbol{E} &= \frac{1}{V}\int_V \hat{\boldsymbol{E}}(x)dx, & \boldsymbol{E}^{(i)} &= \frac{1}{V_i}\int_{V_i} \hat{\boldsymbol{E}}^{(i)}(x)dx \\
\boldsymbol{D} &= \frac{1}{V}\int_V \hat{\boldsymbol{D}}(x)dx, & \boldsymbol{D}^{(i)} &= \frac{1}{V_i}\int_{V_i} \hat{\boldsymbol{D}}^{(i)}(x)dx
\end{aligned} \tag{6}
$$

where $\hat{\boldsymbol{\sigma}}$, $\hat{\boldsymbol{\varepsilon}}$, $\hat{\boldsymbol{E}}$ and $\hat{\boldsymbol{D}}$ are the stress tensor, strain tensor, electric field tensor and electric displacement tensor at random over a representative volume element, respectively; $\boldsymbol{\sigma}$, $\boldsymbol{\varepsilon}$, $\boldsymbol{E}$ and $\boldsymbol{D}$ are the overall volume average stress tensor, strain tensor, electric field tensor and electric displacement tensor of composite material, respectively; $\hat{\boldsymbol{\sigma}}^{(i)}$, $\hat{\boldsymbol{\varepsilon}}^{(i)}$, $\hat{\boldsymbol{E}}^{(i)}$ and $\hat{\boldsymbol{D}}^{(i)}$ are the stress tensor, strain tensor, electric field tensor and electric displacement tensor at random over the constituents of composite material, respectively; and $\boldsymbol{\sigma}^{(i)}$, $\boldsymbol{\varepsilon}^{(i)}$, $\boldsymbol{E}^{(i)}$ and $\boldsymbol{D}^{(i)}$ is the overall volume average stress tensor in their subvolumes $V_i$, respectively.

For the functionally graded piezoelectric cylinder consists of two phases of material A ($i$=1) and B ($i$=2), Eq. (6) reduces to

$$
\begin{aligned}
\boldsymbol{\sigma} &= c(r)\boldsymbol{\sigma}^{(1)} + \left[1 - c(r)\right]\boldsymbol{\sigma}^{(2)} \\
\boldsymbol{\varepsilon} &= c(r)\boldsymbol{\varepsilon}^{(1)} + \left[1 - c(r)\right]\boldsymbol{\varepsilon}^{(2)} \\
D_r &= c(r)D_r^{(1)} + \left(1 - c(r)\right)D_r^{(2)} \\
E_r &= c(r)E_r^{(1)} + \left(1 - c(r)\right)E_r^{(2)}
\end{aligned} \tag{7}
$$

By using the assumption of a uniform strain field within the representative volume element, the components of the strain tensor and electric displacement tensor components are

$$\varepsilon_{\theta}^{(1)} = \varepsilon_{\theta}^{(2)} = \varepsilon_{\theta}, \quad \varepsilon_{r}^{(1)} = \varepsilon_{r}^{(2)} = \varepsilon_{r}, \quad D_{r}^{(1)} = D_{r}^{(2)} = D_{r} \tag{8}$$

For the axisymmetric problem, the strain-displacement relations are

$$\varepsilon_{r} = \frac{du}{dr}, \qquad \varepsilon_{\theta} = \frac{u}{r} \tag{9}$$

where $u$ is the displacement in the radial direction.

Substituting Eq. (5) into $(7)_1$ using Eqs. (8) and (9), the equivalent stresses of the cylinder reduce to

$$\begin{aligned}
\sigma_{r} &= a(r)\frac{u}{r} + b(r)\frac{du}{dr} - f(r)D_{r} \\
\sigma_{\theta} &= d(r)\frac{u}{r} + a(r)\frac{du}{dr} - h(r)D_{r} \\
\sigma_{z} &= g(r)\frac{u}{r} + a(r)\frac{du}{dr} - h(r)D_{r}
\end{aligned} \tag{10}$$

where

$$\begin{aligned}
a(r) &= c(r)\left(C_{13}^{1} + \frac{e_{31}^{1}e_{33}^{1}}{k_{33}^{1}}\right) + \left(1 - c(r)\right)\left(C_{13}^{2} + \frac{e_{31}^{2}e_{33}^{2}}{k_{33}^{2}}\right) \\
b(r) &= c(r)\left(C_{33}^{1} + \frac{\left(e_{33}^{1}\right)^{2}}{k_{33}^{1}}\right) + \left(1 - c(r)\right)\left(C_{33}^{2} + \frac{\left(e_{33}^{2}\right)^{2}}{k_{33}^{2}}\right) \\
d(r) &= c(r)\left(C_{11}^{1} + \frac{\left(e_{31}^{1}\right)^{2}}{k_{33}^{1}}\right) + \left(1 - c(r)\right)\left(C_{11}^{2} + \frac{\left(e_{31}^{2}\right)^{2}}{k_{33}^{2}}\right) \\
g(r) &= c(r)\left(C_{12}^{1} + \frac{\left(e_{31}^{1}\right)^{2}}{k_{33}^{1}}\right) + \left(1 - c(r)\right)\left(C_{12}^{2} + \frac{\left(e_{31}^{2}\right)^{2}}{k_{33}^{2}}\right) \\
f(r) &= c(r)\frac{e_{33}^{1}}{k_{33}^{1}} + \left(1 - c(r)\right)\frac{e_{33}^{2}}{k_{33}^{2}} \\
h(r) &= c(r)\frac{e_{31}^{1}}{k_{33}^{1}} + \left(1 - c(r)\right)\frac{e_{31}^{2}}{k_{33}^{2}}
\end{aligned} \tag{11}$$

By using Eqs. (4) and $(7)_4$, the equivalent radial electric field of the cylinder can be obtained as

$$E_r = k(r)D_r - h(r)\varepsilon_\theta - f(r)\varepsilon_r \tag{12}$$

where $k(r) = \dfrac{c(r)}{k_{33}^1} + \dfrac{1-c(r)}{k_{33}^2}$.

Considering the electric potential $\varphi$, the radial electric field is given by

$$E_r = -\frac{d\varphi}{dr} \tag{13}$$

In the cylindrical coordinate system, equilibrium equation and electrostatic charge equation for axisymmetric deformation, respectively, are

$$\frac{d\sigma_r}{dr} + \frac{\sigma_r - \sigma_\theta}{r} = 0 \tag{14}$$

and

$$\frac{dD_r}{dr} + \frac{D_r}{r} = 0 \tag{15}$$

Eq. (15) can be easily solved to obtain the radial electric displacement as

$$D_r = \frac{C_1}{r} \tag{16}$$

where $C_1$ is integral constant.

Then the radial electric field can be given when substituting Eq. (16) into (12) and using Eq. (9) as

$$E_r = C_1\frac{k(r)}{r} - h(r)\frac{u}{r} - f(r)\frac{du}{dr} \tag{17}$$

And then the electric potential $\varphi$ is solved by comparing Eq. (13) and Eq. (17) as

$$\varphi(r) = \int_a^r \left[\frac{h(r)}{r} - \frac{df(r)}{dr}\right] u dr + f(r)u - C_1\int_a^r \frac{k(r)}{r} dr + C_2 \tag{18}$$

where $C_2$ is integral constant.

From Eq. (18), the specific form of the electric potential $\varphi$ can be obtained once the radial displacement $u$ is given. Then the radial displacement $u$ is given in the following.

Substituting Eq. (10) into (14) and using Eq. (16), the governing differential equation for the radial displacement $u$ is obtained as

$$r\frac{d}{dr}\left[r\left(\phi_1-\phi_2 r^n\right)\frac{du}{dr}\right]+\left(\phi_4 r^n-\phi_3\right)u=C_1\left(\phi_6 r^n-\phi_5\right) \tag{19}$$

where

$$\begin{aligned}
\phi_1 &= c_0\left(C_{33}^1+e_{33}^{1\,2}/k_{33}^1\right)+(1-c_0)\left(C_{33}^2+e_{33}^{2\,2}/k_{33}^2\right)\\
\phi_2 &= c_0 k\left(C_{33}^1+e_{33}^{1\,2}/k_{33}^1-C_{33}^2-e_{33}^{2\,2}/k_{33}^2\right)/b^n\\
\phi_3 &= c_0\left(C_{11}^1+e_{31}^{1\,2}/k_{33}^1\right)+(1-c_0)\left(C_{11}^2+e_{31}^{2\,2}/k_{33}^2\right)\\
\phi_4 &= c_0 k\left[C_{11}^1+e_{31}^{1\,2}/k_{33}^1-C_{11}^2-e_{31}^{2\,2}/k_{33}^2-n\left(C_{13}^1+e_{31}^1 e_{33}^1/k_{33}^1-C_{13}^2-e_{31}^2 e_{33}^2/k_{33}^2\right)\right]/b^n\\
\phi_5 &= c_0\, e_{31}^1/k_{33}^1+(1-c_0)\,e_{31}^2/k_{33}^2\\
\phi_6 &= c_0 k\left[e_{31}^1/k_{33}^1-e_{31}^2/k_{33}^2-n\left(e_{33}^1/k_{33}^1-e_{33}^2/k_{33}^2\right)\right]/b^n
\end{aligned} \tag{20}$$

The value of $\phi_1$ usually not equal to zero due to the volume fracture belong to zero to one, then Eq. (19) can be rewritten as

$$r\frac{d}{dr}\left[r\left(1-\frac{\phi_2}{\phi_1}r^n\right)\frac{du}{dr}\right]+\left(\frac{\phi_4}{\phi_1}r^n-\frac{\phi_3}{\phi_1}\right)u=C_1\left(\frac{\phi_6}{\phi_1}r^n-\frac{\phi_5}{\phi_1}\right) \tag{21}$$

For notational convenience, a new variable $x=\chi(r)=\frac{\phi_2}{\phi_1}r^n$ is introduced here, Eq. (21) then becomes

$$x^2\left(x-1\right)\frac{d^2u}{dx^2}+x\left(2x-1\right)\frac{du}{dx}+\frac{1}{n^2}\left(\frac{\phi_3}{\phi_1}-\frac{\phi_4}{\phi_2}x\right)u=\frac{C_1}{n^2}\left(\frac{\phi_5}{\phi_1}-\frac{\phi_6}{\phi_2}x\right) \tag{22}$$

The solution of Eq. (22) is the sum of the general solution $u_g(r)$ of the corresponding homogeneous ordinary differential equation and a particular solution $u_p(r)$, and $u_g(r)$ can be solved in the range $0<x<1$ as [25]

$$u_g(r)=C_3 x^{\delta/2-1/2}F(\alpha,\beta,\delta;x)+C_4 x^{1/2-\delta/2}F(\alpha-\delta+1,\beta-\delta+1,2-\delta;x) \tag{23}$$

with $C_3$ and $C_4$ are integral constants and $F$ is the hypergeometric function defined by Eq. (11) in Ref. [25] and the constants are

$$\delta=1+2\sqrt{\frac{\phi_3}{n^2\phi_1}},\quad \alpha=\frac{\delta}{2}+\sqrt{\frac{1}{4}+\frac{\phi_4}{n^2\phi_2}},\quad \beta=\delta-\alpha \tag{24}$$

Note that the variable $x=\chi(r)$ mainly lies in the interval $0<x<1$ for different material parameters. Here we do not discuss the solutions of Eq. (22) for other intervals.

For convenience, rewrite $u_p(r)$ as

$$u_g(r) = C_3 P(r) + C_4 Q(r) \tag{25}$$

where $P(r)$, $Q(r)$ and their derivatives $P'(r)$, $Q'(r)$ are

$$P(r) = x^{\delta/2-1/2} F(\alpha, \beta, \delta; x)$$
$$Q(r) = x^{1/2-\delta/2} F(\alpha - \delta + 1, \beta - \delta + 1, 2 - \delta; x)$$
$$P'(r) = nr^{-1} \left[ \frac{\delta - 1}{2} P(r) + \frac{\alpha\beta}{\delta} x^{\frac{\delta}{2}+\frac{1}{2}} F(\alpha + 1, \beta + 1, \delta + 1; x) \right] \tag{26}$$
$$Q'(r) = nr^{-1} \left[ \frac{1-\delta}{2} Q(r) + \frac{(\alpha - \delta + 1)(\beta - \delta + 1)}{2 - \delta} x^{\frac{3}{2}-\frac{\delta}{2}} F(\alpha - \delta + 2, \beta - \delta + 2, 3 - \delta; x) \right]$$

A particular solution can be easily obtained as

$$u_p(r) = C_1 G(r) \tag{27}$$

where $G(r)$ and its derivative $G'(r)$ are

$$G(r) = \frac{\phi_5}{\phi_3} + \frac{\phi_1(\phi_4\phi_5 - \phi_3\phi_6)}{\phi_2\phi_3(\phi_3 - n^2\phi_1)} \left[ x + \sum_{m=2}^{\infty} \frac{C_{\bar{\alpha}+m}^{m-1} C_{\bar{\beta}+m}^{m-1}}{C_{\bar{\delta}+m}^{m-1} C_{-\bar{\delta}+m}^{m-1}} x^m \right]$$
$$G'(r) = \frac{n(\phi_4\phi_5 - \phi_3\phi_6)}{\phi_3(\phi_3 - n^2\phi_1)} r^{n-1} \left[ 1 + \sum_{m=2}^{\infty} m \frac{C_{\bar{\alpha}+m}^{m-1} C_{\bar{\beta}+m}^{m-1}}{C_{\bar{\delta}+m}^{m-1} C_{-\bar{\delta}+m}^{m-1}} x^{m-1} \right] \tag{28}$$

with $C$ is combination symbol and the constants in $G(r)$ and $G'(r)$ are

$$\bar{\delta} = \delta / 2 - 1/2, \quad \bar{\alpha} = \bar{\delta} - \alpha, \quad \bar{\beta} = -\bar{\alpha} - 1 \tag{29}$$

Now the solution of Eq. (22) is the sum of Eqs. (25) and (27), namely

$$u(r) = C_3 P(r) + C_4 Q(r) + C_1 G(r) \tag{30}$$

Then the stress components can be obtained substituting Eq. (30) into (10) as

$$\sigma_r = a(r)\left[C_3 P(r) + C_4 Q(r) + C_1 G(r)\right]/r + b(r)\left[C_3 P'(r) + C_4 Q'(r) + C_1 G'(r)\right] - C_1 f(r)/r$$
$$\sigma_\theta = d(r)\left[C_3 P(r) + C_4 Q(r) + C_1 G(r)\right]/r + a(r)\left[C_3 P'(r) + C_4 Q'(r) + C_1 G'(r)\right] - C_1 h(r)/r$$
$$\sigma_z = g(r)\left[C_3 P(r) + C_4 Q(r) + C_1 G(r)\right]/r + a(r)\left[C_3 P'(r) + C_4 Q'(r) + C_1 G'(r)\right] - C_1 h(r)/r$$

(31)

And then the electric potential $\varphi$ is given substituting Eq. (30) into (18) as

$$\varphi(r) = C_3\left[A(r) + P(r) f(r)\right] + C_4\left[B(r) + Q(r) f(r)\right]$$
$$+ C_1 \left[ C(r) + G(r) f(r) - \left( \frac{c_0}{k_{33}^1} + \frac{1 - c_0}{k_{33}^2} \right) \ln r + \frac{c_0 k(k_{33}^2 - k_{33}^1)}{nb^n k_{33}^1 k_{33}^2} r^n \right] + C_2 \tag{32}$$

where

$$
\begin{aligned}
A(r) &= \int_a^r \left[ \frac{h(r)}{r} - \frac{df(r)}{dr} \right] P(r)dr \\
&= A_1 r^{\frac{n\delta}{2}-\frac{n}{2}} \left[ \sum_{m=1}^{\infty} \frac{C_{\alpha+m-1}^{m} C_{\beta+m-1}^{m}}{C_{\delta+m-1}^{m}} \frac{x^m}{m+\frac{\delta}{2}-\frac{1}{2}} + \frac{1}{\frac{\delta}{2}-\frac{1}{2}} \right] + A_2 r^{\frac{n\delta}{2}+\frac{n}{2}} \left[ \sum_{m=1}^{\infty} \frac{C_{\alpha+m-1}^{m} C_{\beta+m-1}^{m}}{C_{\delta+m-1}^{m}} \frac{x^m}{m+\frac{\delta}{2}+\frac{1}{2}} + \frac{1}{\frac{\delta}{2}+\frac{1}{2}} \right] \\
&\quad - A_1 a^{\frac{n\delta}{2}-\frac{n}{2}} \left[ \sum_{m=1}^{\infty} \frac{C_{\alpha+m-1}^{m} C_{\beta+m-1}^{m}}{C_{\delta+m-1}^{m}} \frac{\chi(a)^m}{m+\frac{\delta}{2}-\frac{1}{2}} + \frac{1}{\frac{\delta}{2}-\frac{1}{2}} \right] - A_2 a^{\frac{n\delta}{2}+\frac{n}{2}} \left[ \sum_{m=1}^{\infty} \frac{C_{\alpha+m-1}^{m} C_{\beta+m-1}^{m}}{C_{\delta+m-1}^{m}} \frac{\chi(a)^m}{m+\frac{\delta}{2}+\frac{1}{2}} + \frac{1}{\frac{\delta}{2}+\frac{1}{2}} \right] \\
B(r) &= \int_a^r \left[ \frac{h(r)}{r} - \frac{df(r)}{dr} \right] Q(r)dr \\
&= A_3 r^{\frac{n}{2}-\frac{n\delta}{2}} \left[ \sum_{m=1}^{\infty} \frac{C_{\alpha-\delta+m}^{m} C_{\beta-\delta+m}^{m}}{C_{1-\delta+m}^{m}} \frac{x^m}{m+\frac{1}{2}-\frac{\delta}{2}} + \frac{1}{\frac{1}{2}-\frac{\delta}{2}} \right] + A_4 r^{\frac{3n}{2}-\frac{n\delta}{2}} \left[ \sum_{m=1}^{\infty} \frac{C_{\alpha-\delta+m}^{m} C_{\beta-\delta+m}^{m}}{C_{1-\delta+m}^{m}} \frac{x^m}{m+\frac{3}{2}-\frac{\delta}{2}} + \frac{1}{\frac{3}{2}-\frac{\delta}{2}} \right] \\
&\quad - A_3 a^{\frac{n}{2}-\frac{n\delta}{2}} \left[ \sum_{m=1}^{\infty} \frac{C_{\alpha-\delta+m}^{m} C_{\beta-\delta+m}^{m}}{C_{1-\delta+m}^{m}} \frac{\chi(a)^m}{m+\frac{1}{2}-\frac{\delta}{2}} + \frac{1}{\frac{1}{2}-\frac{\delta}{2}} \right] - A_4 a^{\frac{3n}{2}-\frac{n\delta}{2}} \left[ \sum_{m=1}^{\infty} \frac{C_{\alpha-\delta+m}^{m} C_{\beta-\delta+m}^{m}}{C_{1-\delta+m}^{m}} \frac{\chi(a)^m}{m+\frac{3}{2}-\frac{\delta}{2}} + \frac{1}{\frac{3}{2}-\frac{\delta}{2}} \right]
\end{aligned}
$$

$$
\begin{aligned}
C(r) &= \int_a^r \left[ \frac{h(r)}{r} - \frac{df(r)}{dr} \right] G(r)dr \\
&= A_5 \left\{ \frac{\phi_5}{\phi_3} \ln\frac{r}{a} + \frac{(\phi_4\phi_5 - \phi_3\phi_6)(r^n - a^n)}{n\phi_3(\phi_3 - n^2\phi_1)} + \frac{\phi_1(\phi_4\phi_5 - \phi_3\phi_6)}{n\phi_2\phi_3(\phi_3 - n^2\phi_1)} \sum_{m=2}^{\infty} \frac{C_{\bar{\alpha}+m}^{m-1} C_{\bar{\beta}+m}^{m-1}}{C_{\bar{\delta}+m}^{m-1} C_{-\bar{\delta}+m}^{m-1}} \frac{\left[ x^m - \chi(a)^m \right]}{m} \right\} \\
&+ A_6 \left\{ \frac{\phi_5(r^n - a^n)}{n\phi_3} + \frac{(\phi_4\phi_5 - \phi_3\phi_6)(r^{2n} - a^{2n})}{2n\phi_3(\phi_3 - n^2\phi_1)} + \frac{\phi_1(\phi_4\phi_5 - \phi_3\phi_6)}{n\phi_2\phi_3(\phi_3 - n^2\phi_1)} \sum_{m=2}^{\infty} \frac{C_{\bar{\alpha}+m}^{m-1} C_{\bar{\beta}+m}^{m-1}}{C_{\bar{\delta}+m}^{m-1} C_{-\bar{\delta}+m}^{m-1}} \frac{\left[ x^m r^n - \chi(a)^m a^n \right]}{m+1} \right\}
\end{aligned} \tag{33}
$$

and the constants in $A(r)$, $B(r)$ and $C(r)$ are

$$
\begin{aligned}
&A_5 = \frac{c_0 e_{31}^1}{k_{33}^1} + \frac{(1-c_0)e_{31}^2}{k_{33}^2}, \qquad A_6 = \frac{c_0 k}{b^n}\left[ \frac{e_{31}^2}{k_{33}^2} - \frac{e_{31}^1}{k_{33}^1} - n\left( \frac{e_{33}^2}{k_{33}^2} - \frac{e_{33}^1}{k_{33}^1} \right) \right], \\
&A_1 = \frac{A_5}{n}\left(\frac{\phi_2}{\phi_1}\right)^{\frac{\delta}{2}-\frac{1}{2}}, \quad A_2 = \frac{A_6}{n}\left(\frac{\phi_2}{\phi_1}\right)^{\frac{\delta}{2}-\frac{1}{2}}, \quad A_3 = \frac{A_5}{n}\left(\frac{\phi_2}{\phi_1}\right)^{\frac{1}{2}-\frac{\delta}{2}}, \quad A_4 = \frac{A_6}{n}\left(\frac{\phi_2}{\phi_1}\right)^{\frac{1}{2}-\frac{\delta}{2}}
\end{aligned} \tag{34}
$$

Until now, the radial displacement, the stress components and the electric potential are all obtained. In the following, the integral constants $C_1$, $C_2$, $C_3$ and $C_4$ are solved using boundary conditions.

With the stress boundary conditions $\sigma_r\big|_{r=a} = -p_a$ and $\sigma_r\big|_{r=b} = -p_b$, the relationships between $C_3$, $C_4$ and $C_1$ are

$$
\begin{aligned}
C_3 &= C_{11}C_1 + C_{12} \\
C_4 &= C_{21}C_1 + C_{22}
\end{aligned} \tag{35}
$$

where

$$
\begin{aligned}
C_{11} &= \Big\{\Big[ f(a)\big/a - a(a)G(a)\big/a - b(a)G'(a) \Big]\big[a(b)Q(b)/b + b(b)Q'(b)\big] \\
&\quad - \Big[ f(b)\big/b - a(b)G(b)\big/b - b(b)G'(b) \Big]\big[a(a)Q(a)/a + b(a)Q'(a)\big]\Big\}\Big/C_0 \\
C_{12} &= \big\{ p_b\big[a(a)Q(a)/a + b(a)Q'(a)\big] - p_a\big[a(b)Q(b)/b + b(b)Q'(b)\big]\big\}\big/C_0 \\
C_{21} &= \Big\{\Big[ f(b)\big/b - a(b)G(b)\big/b - b(b)G'(b) \Big]\big[a(a)P(a)/a + b(a)P'(a)\big] \\
&\quad - \Big[ f(a)\big/a - a(a)G(a)\big/a - b(a)G'(a) \Big]\big[a(b)P(b)/b + b(b)P'(b)\big]\Big\}\Big/C_0 \\
C_{22} &= \big\{ p_a\big[a(b)P(b)/b + b(b)P'(b)\big] - p_b\big[a(a)P(a)/a + b(a)P'(a)\big]\big\}\big/C_0
\end{aligned} \tag{36}
$$

and

$$
\begin{aligned}
C_0 &= \big[a(a)P(a)/a + b(a)P'(a)\big]\big[a(b)Q(b)/b + b(b)Q'(b)\big] \\
&\quad - \big[a(a)Q(a)/a + b(a)Q'(a)\big]\big[a(b)P(b)/b + b(b)P'(b)\big]
\end{aligned} \tag{37}
$$

Then the potential electric boundary conditions $\varphi\big|_{r=a} = \varphi_a$ and $\varphi\big|_{r=b} = \varphi_b$ are used for solving $C_1$ and $C_2$, and we have

$$
\begin{aligned}
C_1 &= \big\{\varphi_b - \varphi_a - C_{12}\big[A(b) + P(b)f(b) - P(a)f(a)\big] \\
&\quad - C_{22}\big[B(b) + Q(b)f(b) - Q(a)f(a)\big]\big\}\big/\bar{C}_0 \\
C_2 &= \varphi_a - \big(C_{12} + C_1C_{11}\big)P(a)f(a) - \big(C_{22} + C_1C_{21}\big)Q(a)f(a) \\
&\quad - C_1\left[ G(a)f(a) - \left(\frac{c_0}{k_{33}^1} + \frac{1-c_0}{k_{33}^2}\right)\ln a + \frac{c_0ka^n(k_{33}^2 - k_{33}^1)}{nb^nk_{33}^1k_{33}^2} \right]
\end{aligned} \tag{38}
$$

and

$$
\begin{aligned}
\bar{C}_0 &= C_{11}\big[A(b) + P(b)f(b) - P(a)f(a)\big] + C_{21}\big[Q(b)f(b) \\
&\quad + B(b) - Q(a)f(a)\big] + C(b) + G(b)f(b) - G(a)f(a) \\
&\quad - \left(\frac{c_0}{k_{33}^1} + \frac{1-c_0}{k_{33}^2}\right)\ln\frac{b}{a} + \frac{c_0k(b^n - a^n)(k_{33}^2 - k_{33}^1)}{na^nk_{33}^1k_{33}^2}
\end{aligned} \tag{39}
$$

In particular, the cylinder will be homogeneous piezoelectric material if the volume fraction is zero or one. For example, $c_0$ =0 means the cylinder totally consists of material B; $c_0$ =1 and $k$=0 means the cylinder totally consists of material A. Then Eq. (20) reduce to

$$
r^2\frac{d^2u}{dr^2} + r\frac{du}{dr} - \frac{\phi_3}{\phi_1}u = -C_1\frac{\phi_5}{\phi_1} \tag{40}
$$

where the constants in Eq. (41) are list below and $i$=1 and 2 denote materials A and B, respectively.

$$\begin{aligned} \phi_1 &= C_{33}^i + {e_{33}^i}^2 / k_{33}^i \\ \phi_3 &= C_{11}^i + {e_{31}^i}^2 / k_{33}^i \\ \phi_5 &= e_{31}^i / k_{33}^i \end{aligned} \tag{41}$$

Also, the solutions of the radial displacement, the stresses, and the electric potential can simplify from Eqs. (30), (31) and (32). Since the results are tedious and nothing but a routine work, they are not presented in detail here.

**3. Results and discussion**

In this numerical part, the non-dimensional expressions for the radial coordinate, the inner radius, the electric potential, the stress and the radial displacement are defined as $\bar{r} = r/b$, $\bar{a} = a/b$, $\bar{\varphi} = \varphi(r)/\varphi_a$, $\bar{\sigma}_{ij} = \sigma_{ij}(r)/p_a$ and $\bar{u} = u(r)C_{11}^1/(bp_a)$, respectively; where $a$ and $b$ are the inner and outer radii, respectively, $p_a$ is the internal pressure and $C_{11}^1$ is one of elastic modulus of material A. The inner radius is taken as $\bar{a} = 0.7$, which is reasonable for a thick-walled cylinder. Here, the internal pressure $p_a$ is taken as 100 KPa and the internal electric potential $\varphi_a$ is taken as 100V. Elastic, piezoelectric and dielectric properties of homogeneous PZT4 [26] and PVDF [27] are listed in table 1.

The volume fraction $c(\bar{r}) = c_0\left[1 - k(\bar{r})^n\right] \in [0,1]$ given in Eq. (1) can describe a wide range profiles by choosing proper material parameters $c_0$, $k$ and $n$. As shown in Fig. 2, we can see the evolution of $c(\bar{r})$ with different parameters in the radial direction. In the case of $c_0$=1, $k$=0, $c(\bar{r})$ equal to one and the cylinder is homogeneously and fully constituted by PZT4. In the functionally graded piezoelectric cylinder simulation, the parameters in volume fraction are chosen as $c_0$=1, $k$=1, throughout.

To illustrate the difference between functionally graded and homogeneous piezoelectric material, the elastic results of homogeneous PZT4 are also displayed in figures. In the following simulation, two representative boundary conditions are applied.

**Example 1.** Consider the mechanical and electric potential boundary conditions of functionally graded piezoelectric cylinder are, respectively, taken as

$$p_a=100\text{KPa}, \quad p_b=0, \quad \varphi_a=0, \quad \varphi_b=0 \tag{42}$$

As shown in Fig. 3, the evolution of radial displacement is plotted with different parameter $n$. The maximum and minimum values of the radial displacement appear at the inner and outer radii, respectively. Comparing with the homogeneous PZT4, we know the values of radial displacement for different parameter $n$ of functionally graded piezoelectric material are all bigger than homogeneous PZT4. Moreover, the larger the parameter $n$ is, the closer to the curve of homogeneous PZT4. This phenomenon is mainly attributed to the change in volume fraction. We can see the curve of $n$=5 is closer to the curve of homogeneous PZT4 ($c_0$=1, $k$=0). Also, the figure shows that the larger the value of parameter $n$ is, the less the relative change occurs in radial displacement.

The radial, circumferential and axial stresses are displayed in Figs. 4, 5 and 6, respectively. Compared to the effects on the radial displacement, the effects of parameter $n$ on the stresses are not obvious. Although parameter $n$ has little effect on stresses of functionally graded piezoelectric cylinder, it is different from homogeneous piezoelectric material. The stress boundary condition $\bar{\sigma}_r\big|_{\bar{r}=\bar{a}}=-1$ and $\bar{\sigma}_r\big|_{\bar{r}=\bar{b}}=0$ are satisfied and the radial stress of homogeneous piezoelectric material looks closer to linearity (see Fig. 4). Fig. 5 shows that the circumferential stress monotonically decreases from the inner radius to the outer radius, and this also applies to the axial stress in Fig. 6.

In this work, we also discuss the difference $\bar{\sigma}_\theta - \bar{\sigma}_r$ between the circumferential stress $\bar{\sigma}_\theta$ and the radial stress $\bar{\sigma}_r$. Fig. 7 shows the evolution of $\bar{\sigma}_\theta - \bar{\sigma}_r$ in the radial direction with different parameter *n*, moreover, the curve $\bar{\sigma}_\theta - \bar{\sigma}_r$ of homogeneous PZT4 is plotted as well to illustrate the difference between FGPM and homogeneous material. We can see that $\bar{\sigma}_\theta - \bar{\sigma}_r$ of homogeneous PZT4 is gentler than that of FGPM, which means FGPM is easier to enter plasticity than that of homogeneous material in this boundary condition.

Fig. 8 depicts the change in electric potential with parameter *n*. It is not hard to see that the electric potential obeys the electric boundary conditions $\bar{\varphi}\big|_{\bar{r}=\bar{a}} = 0$ and $\bar{\varphi}\big|_{\bar{r}=\bar{b}} = 0$. The maximum value in magnitude of the electric potential is located near in the middle of the cylinder. This maximum value increases with an increasing parameter *n*, which means if you want to get a large or small voltage in the middle position, you should choose the right value of parameter *n* accordingly. Furthermore, although the maximum value of the electric potential occurs at an internal position, the position slightly shifts with parameter *n* varying. The larger value of parameter *n* is, the closer the peak is to the outside of the cylinder.

**Example 2.** Consider the mechanical and electric potential boundary conditions of functionally graded piezoelectric cylinder are, respectively, taken as

$$p_a=0, \quad p_b=0, \quad \varphi_a=100\text{V}, \quad \varphi_b=0 \tag{43}$$

Fig. 9 shows the radial displacement as a function of radius. It can be seen from the Fig. 9 that the displacement of the uniform piezoelectric material gradually increases from the inside to the outside of the cylinder, while the functionally graded piezoelectric cylinder increases first and then decreases along the radial direction. Moreover, the value of radial displacement for FGPM is much smaller than that of uniform material. As parameter *n* increases, the magnitude of the radial displacement increases.

Figs. 10, 11 and 12 show the radial stress, circumferential stress, and axial stress as a function of radius in the radial direction. The value of these three stress components are less at the electric potential boundary condition than the stress boundary condition. For the radial stress, it is easy to see the stress boundary condition $\bar{\sigma}_r\big|_{\bar{r}=\bar{a}}=0$ and $\bar{\sigma}_r\big|_{\bar{r}=\bar{b}}=0$ satisfy Eq. (43) whether it is FGPM or homogeneous piezoelectric material. The large value of radial stress is found in FGPM when compared with the homogeneous piezoelectric material. The maximum value of the radial stress is approximately presented in the middle of the cylinder and varies with different parameter *n*-value locations. And as *n* increases, the maximum value of the radial stress shifts to the inside of the cylinder. As shown in Fig. 11, the circumferential stress gradually increases along the radial direction. An interesting phenomenon can be seen here is the value of circumferential stress is almost the same at the outer radius. The circumferential stress of the uniform material changes greatly along the radial direction, and the difference between inside and outside reaches 34.15. The change of axial stress (Fig. 12) and the circumferential stress are similar as the radius increases. Also, an interesting phenomenon can be found that the value of axial stress is basically the same on the outer surface of the cylinder.

In this example, we also discuss the difference $\bar{\sigma}_\theta-\bar{\sigma}_r$ between the circumferential stress $\bar{\sigma}_\theta$ and the radial stress $\bar{\sigma}_r$. Fig. 13 shows the evolution of $\bar{\sigma}_\theta-\bar{\sigma}_r$ in the radial direction with different parameter *n* for FGPM and Homogeneous material. We can see that the curve of $\bar{\sigma}_\theta-\bar{\sigma}_r$ is steep for homogeneous PZT4 but gentle for FGPM. So the designer can control the value of parameter *n* to make the different positions of the cylinder reach plastic at the same time, which is very helpful for engineering practice.

Fig. 14 plots the variation of electric potential with parameter *n*. Clearly, the electric boundary conditions $\bar{\varphi}\big|_{\bar{r}=\bar{a}}=1$ and $\bar{\varphi}\big|_{\bar{r}=\bar{b}}=0$ are in accordance with Eq. (43). When the value of parameter *n* is small, the electric potential change is close to a straight line; and as the value of parameter *n* increases, the nonlinear trend of the electric potential becomes more and more obvious. The electric potential curve of the homogeneous piezoelectric material is close to the case of *n*=1.5 of functionally graded piezoelectric material.

## 4. Conclusions

The exact solutions of the radial displacement, the stress components and the electric potential of a functionally graded piezoelectric hollow thick-walled cylinder under axisymmetric mechanical and electric loads are conducted in current work. The present method can avoid the assumption of the distribution regularities of unknown overall material parameters that appear in existing papers. Furthermore, the solutions in this work are suitable for more complex changes due to that the form of the volume fraction is three variable parameters. Numerical simulation compared the solutions of the functionally graded piezoelectric cylinder and homogeneous piezoelectric cylinder and some conclusions can be obtained as follows: (a) For stress-dominated boundary conditions: the effects of parameter *n* on the stresses are not obvious but are huge in radial displacement and electric potential. Furthermore, the maximum value of the electric potential occurs at an internal position and the position slightly shifts as parameter *n* varies. The larger the value of parameter *n* is, the closer the peak is to the outside of the cylinder. (b) For electric potential-dominated boundary conditions: the value of radial displacement for FGPM is much smaller than that of the uniform material. The maximum value of the radial stress is approximately presented in the middle of the cylinder and varies with different parameter *n*-value locations. And as *n* increases, the maximum value of the radial stress shifts to the inside of the cylinder. Moreover, the difference between the circumferential stress and the radial stress is discussed and the designer can choose suitable parameter *n* to make the different positions of the cylinder reach plastic at the same time.

**Acknowledgments**

The authors acknowledge the financial support of National Natural Science Foundation of China (No. 11772041).

## Tables

**Table 1.**

Elastic, piezoelectric and dielectric properties of homogeneous material A (PZT4) and material B (PVDF).

| Moduli | PZT4 ($i$=1) | PVDF ($i$=2) |
|---|---|---|
| $C_{11}^{i}$ (GPa) | 139 | 3 |
| $C_{22}^{i}$ | 139 | 3 |
| $C_{33}^{i}$ | 115 | 3 |
| $C_{12}^{i}$ | 77.8 | 1.5 |
| $C_{13}^{i}$ | 74.3 | 1.5 |
| $C_{23}^{i}$ | 74.3 | 1.5 |
| $C_{44}^{i}$ | 25.6 | 0.75 |
| $C_{55}^{i}$ | 25.6 | 0.75 |
| $C_{66}^{i}$ | 30.6 | 0.75 |
| $e_{31}^{i}$ (C/m$^2$) | -5.2 | -0.0015 |
| $e_{32}^{i}$ | -5.2 | 0.0285 |
| $e_{33}^{i}$ | 15.1 | -0.051 |
| $e_{24}^{i}$ | 12.7 | — |
| $e_{15}^{i}$ | 12.7 | — |
| $k_{11}^{i}$ (F/m) | 3.27e-9 | 0.1062e-9 |
| $k_{22}^{i}$ | 3.27e-9 | 0.1062e-9 |
| $k_{33}^{i}$ | 5.62e-9 | 0.1062e-9 |

## Figure captions

**Fig. 1.** Schematic diagram of a long thick-walled functionally graded piezoelectric cylinder subjected to mechanical and electric loadings.

**Fig. 2.** Evolution of volume fraction with different parameters

**Fig. 3.** Evolution of the radial displacement with different parameter *n*

**Fig. 4.** Evolution of the radial stress with different parameter *n*

**Fig. 5.** Evolution of the circumferential stress with different parameter *n*

**Fig. 6.** Evolution of the axial stress with different parameter *n*

**Fig. 7.** Evolution of the difference between the circumferential stress and the radial stress with different parameter *n*

**Fig. 8.** Evolution of the electric potential with different parameter *n*

**Fig. 9.** Evolution of the radial displacement with different parameter *n*

**Fig. 10.** Evolution of the radial stress with different parameter *n*

**Fig. 11.** Evolution of the circumferential stress with different parameter *n*

**Fig. 12.** Evolution of the axial stress with different parameter *n*

**Fig. 13.** Evolution of the difference between the circumferential stress and the radial stress with different parameter *n*

**Fig. 14.** Evolution of the electric potential with different parameter *n*

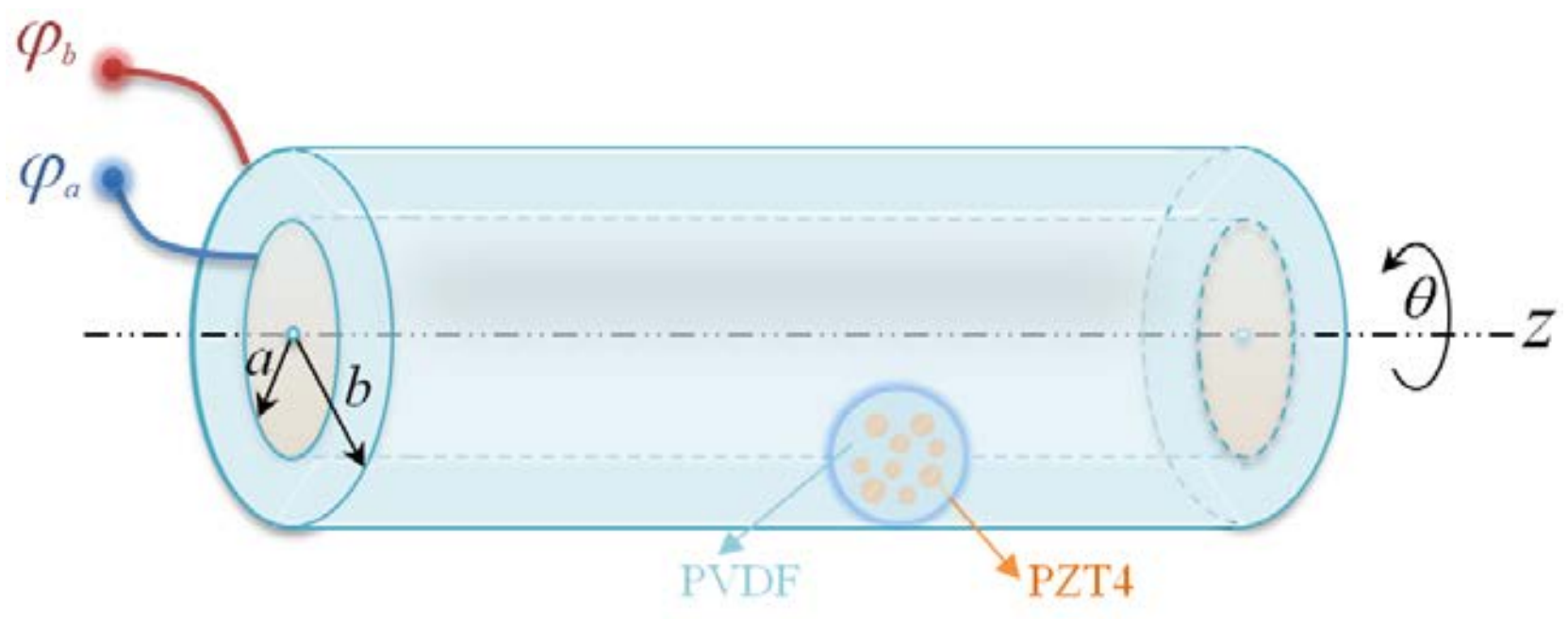

**Fig. 1.** Schematic diagram of a long thick-walled functionally graded piezoelectric cylinder subjected to mechanical and electric loadings.

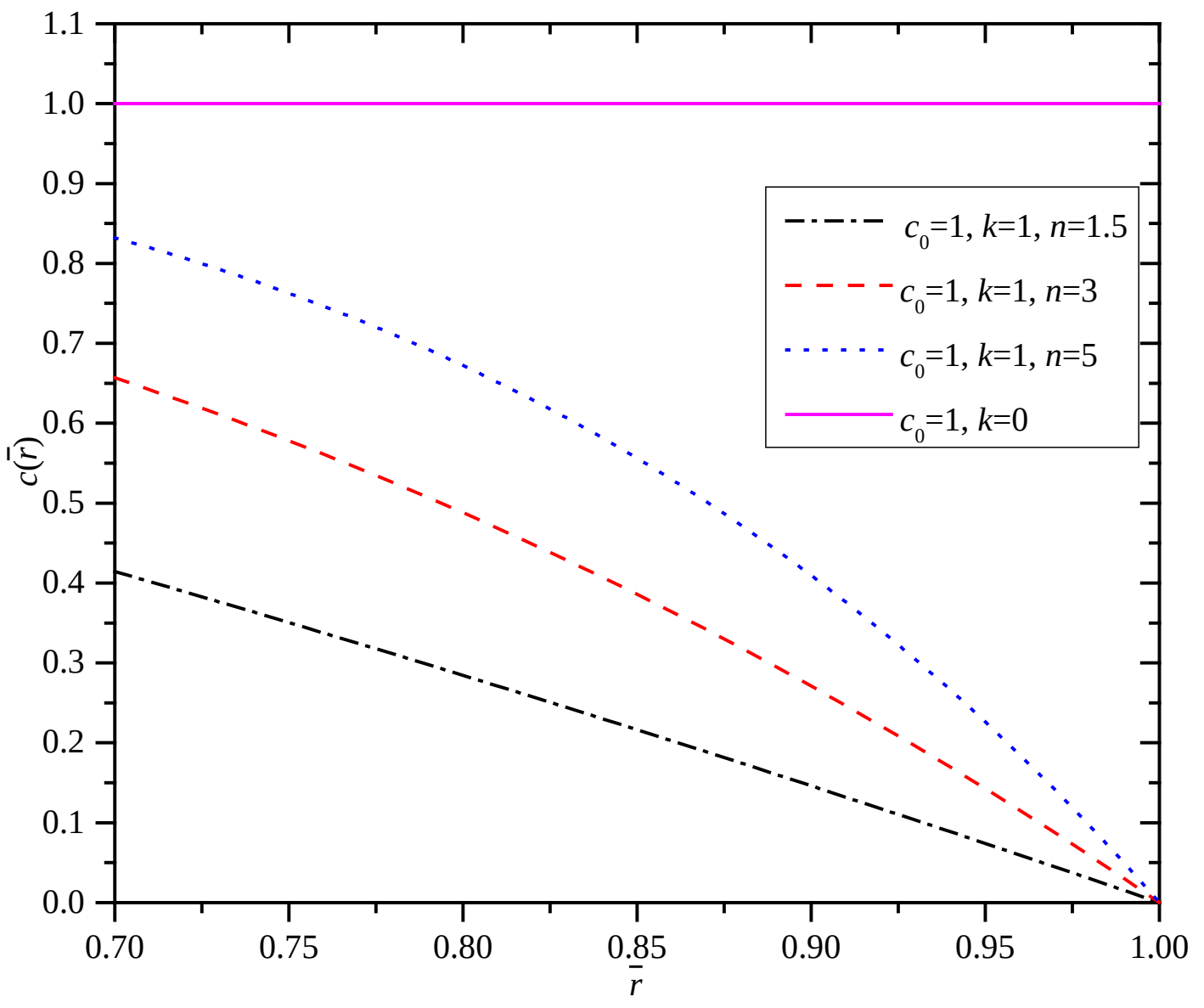

**Fig. 2.** Evolution of volume fraction with different parameters

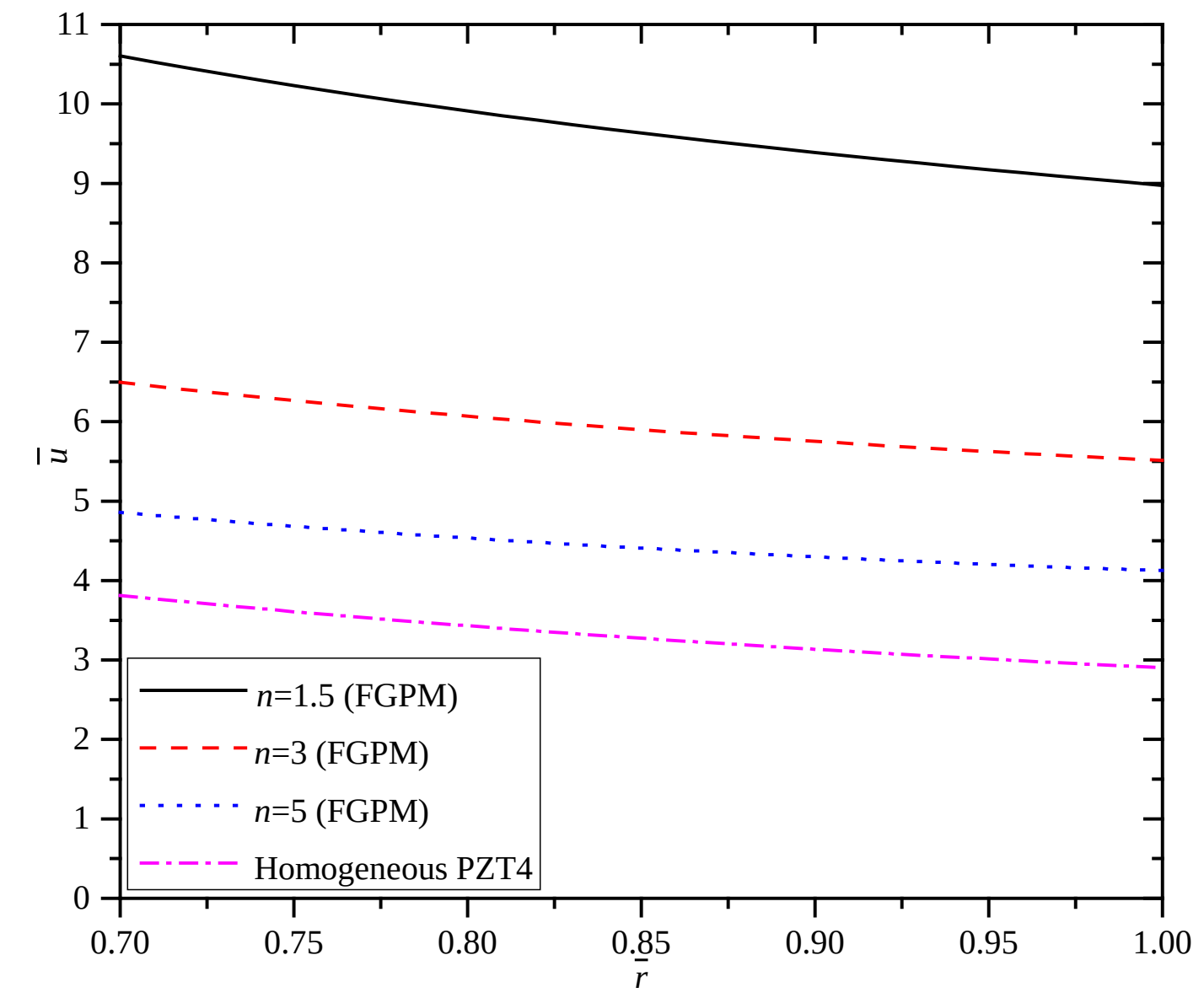

**Fig. 3.** Evolution of the radial displacement with different parameter *n*

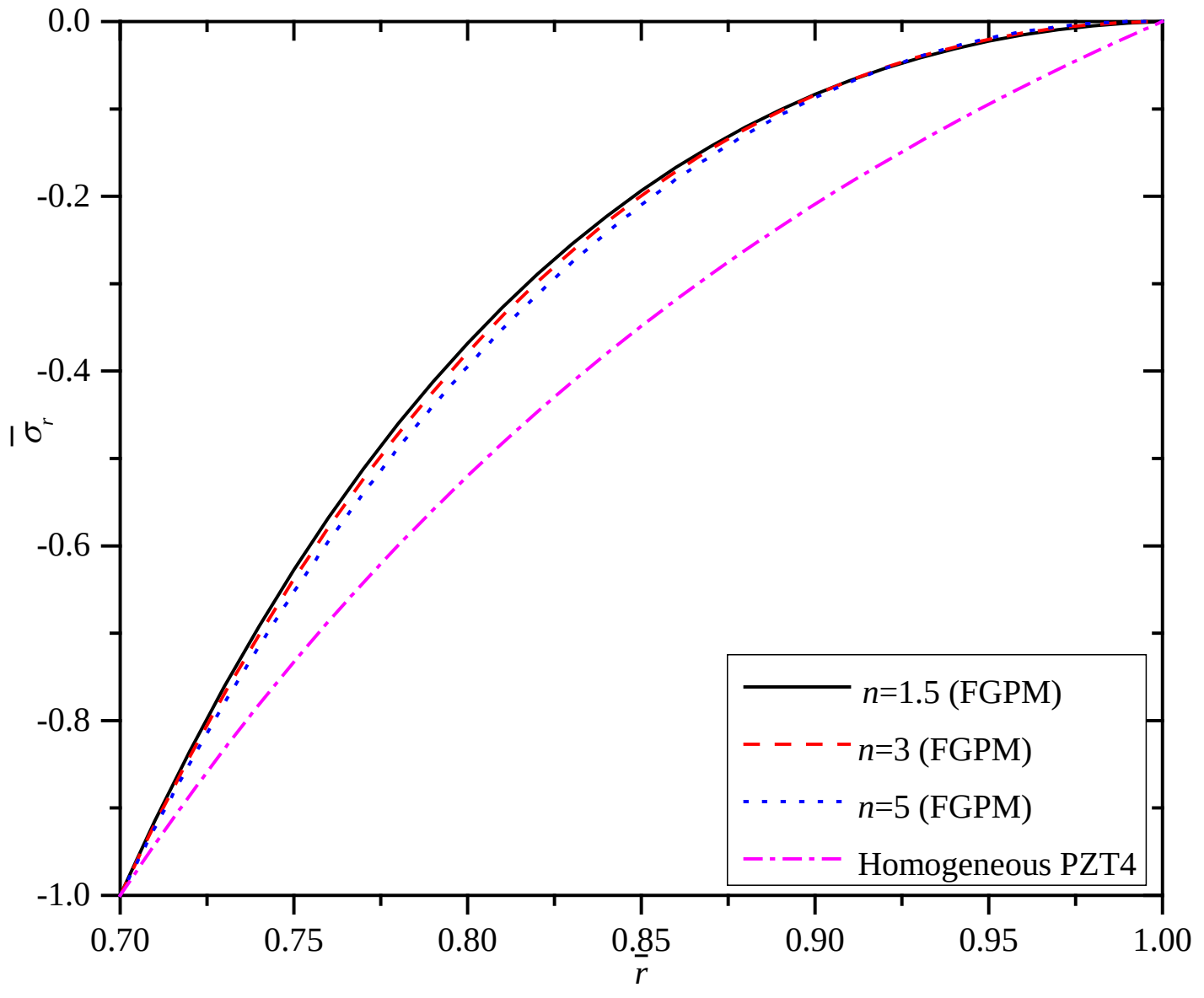

**Fig. 4.** Evolution of the radial stress with different parameter *n*

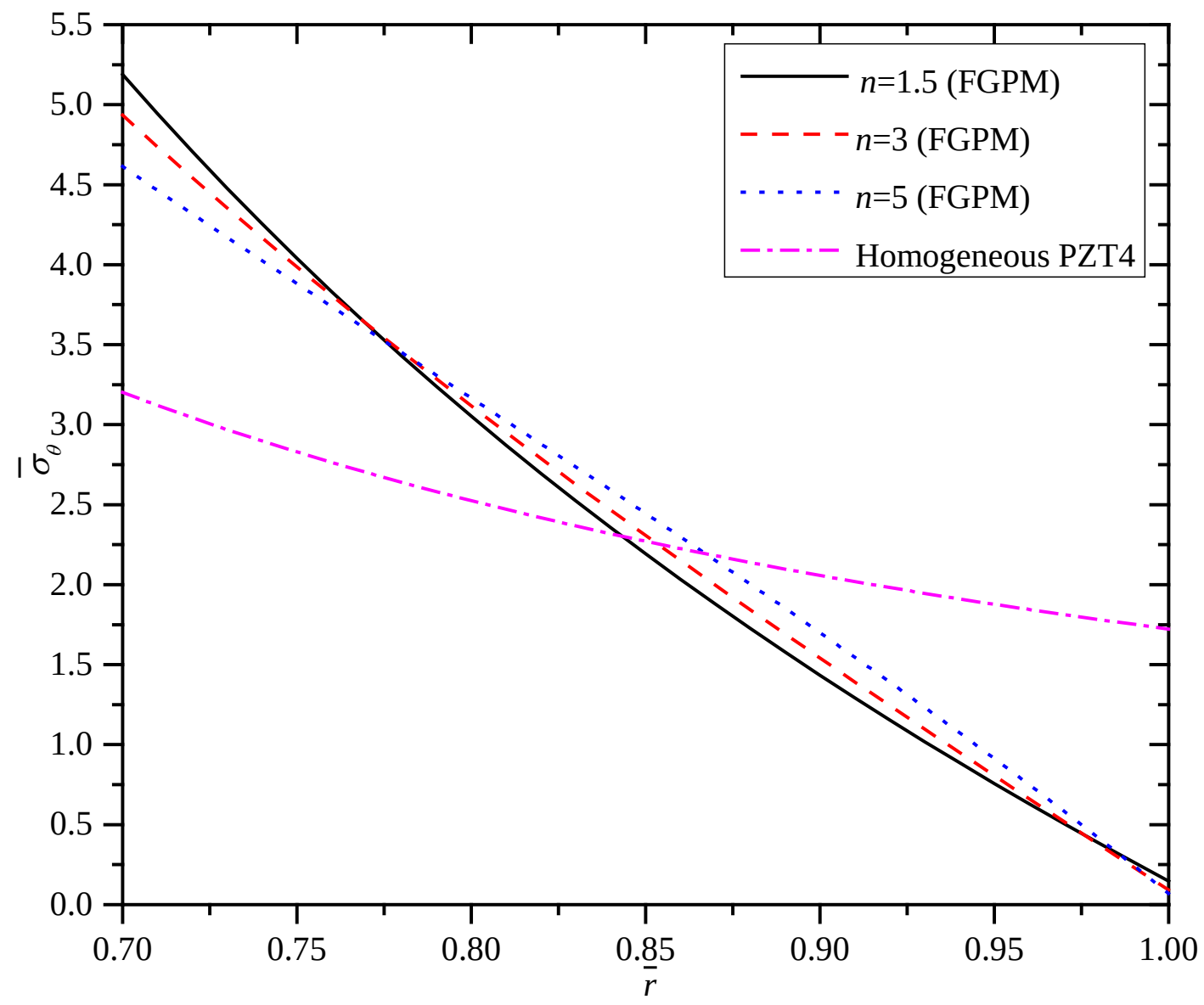

**Fig. 5.** Evolution of the circumferential stress with different parameter *n*

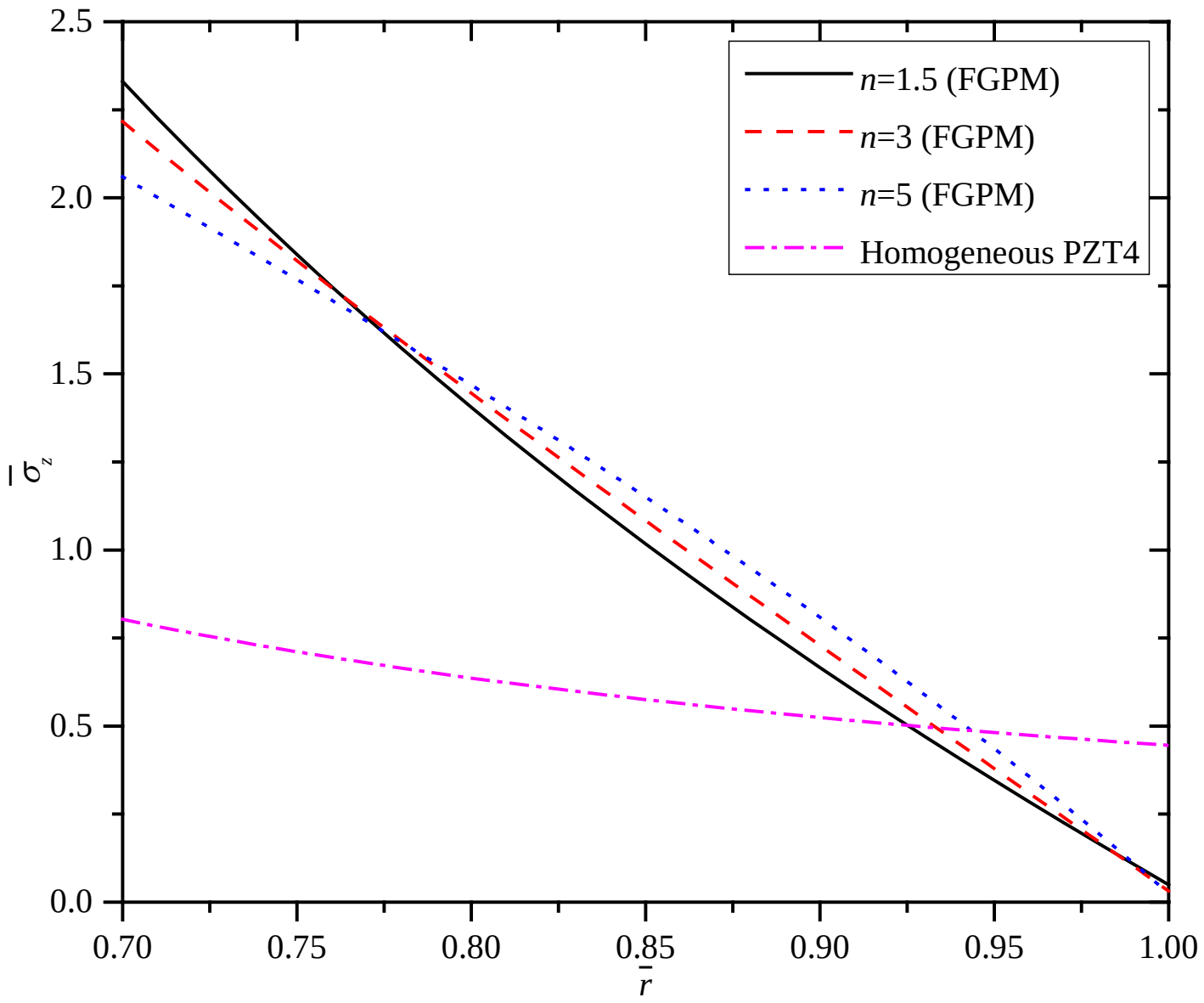

**Fig. 6.** Evolution of the axial stress with different parameter *n*

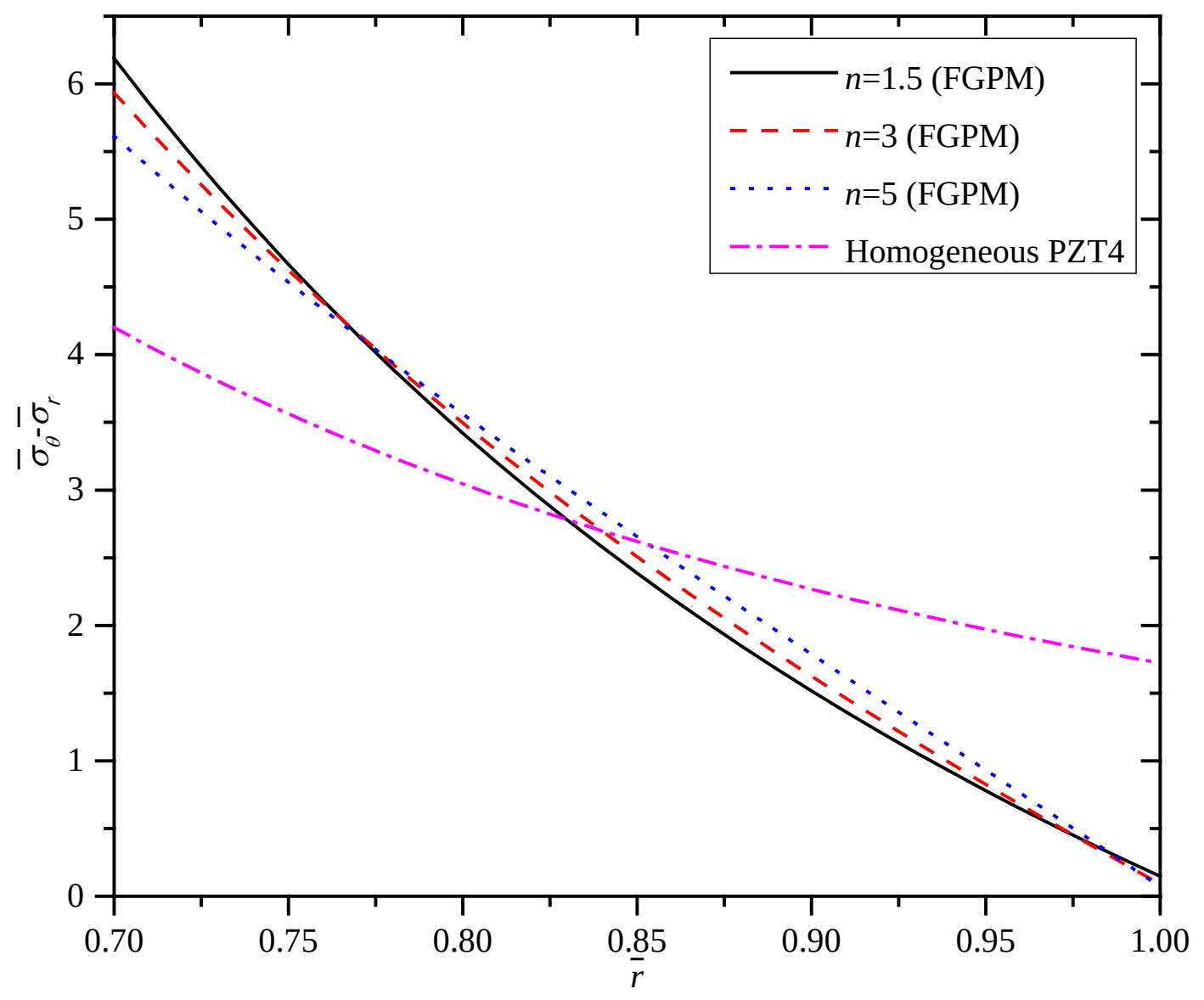

**Fig. 7.** Evolution of the difference between the circumferential stress and the radial stress with different parameter *n*

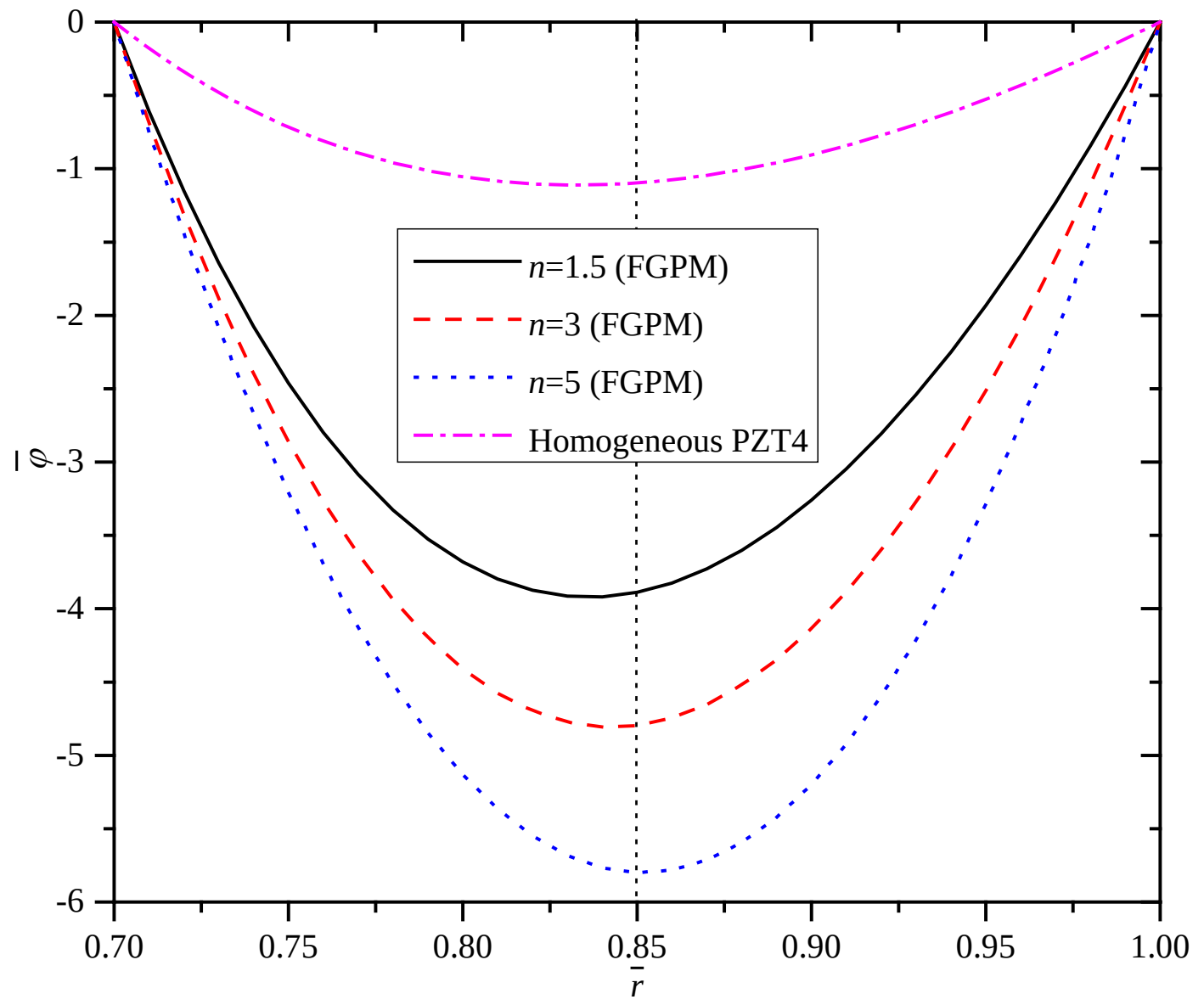

**Fig. 8.** Evolution of the electric potential with different parameter *n*

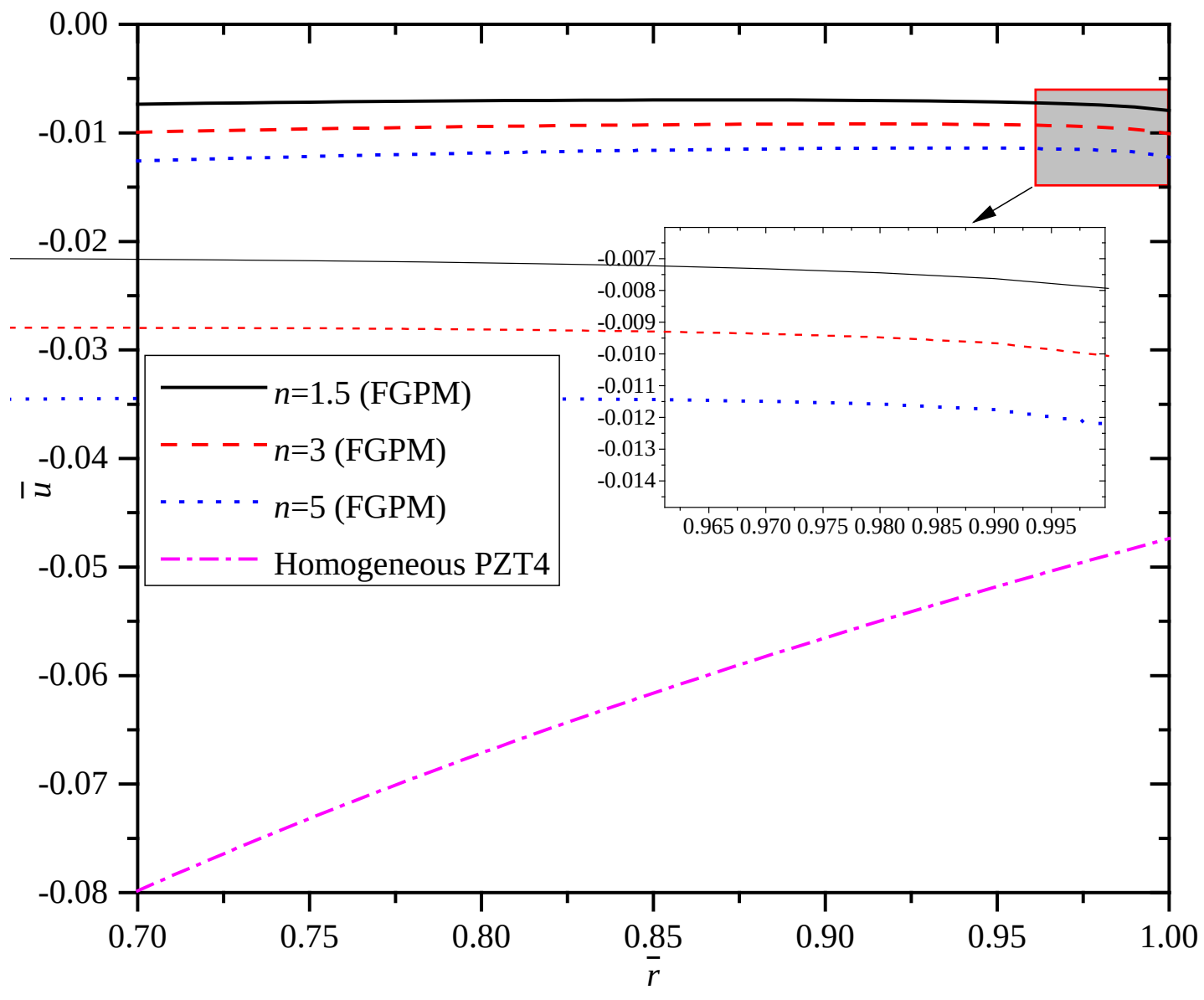

**Fig. 9.** Evolution of the radial displacement with different parameter *n*

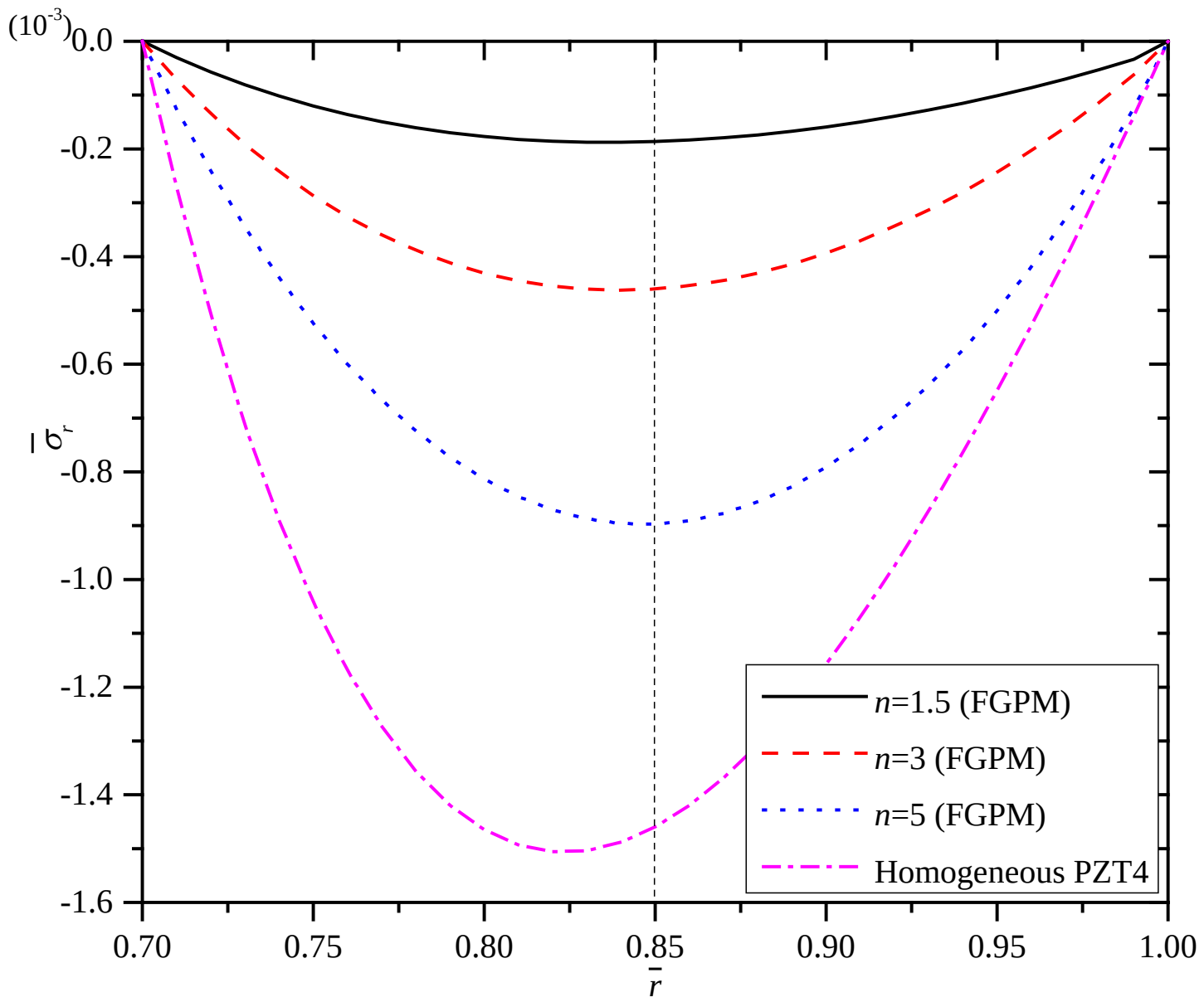

**Fig. 10.** Evolution of the radial stress with different parameter *n*

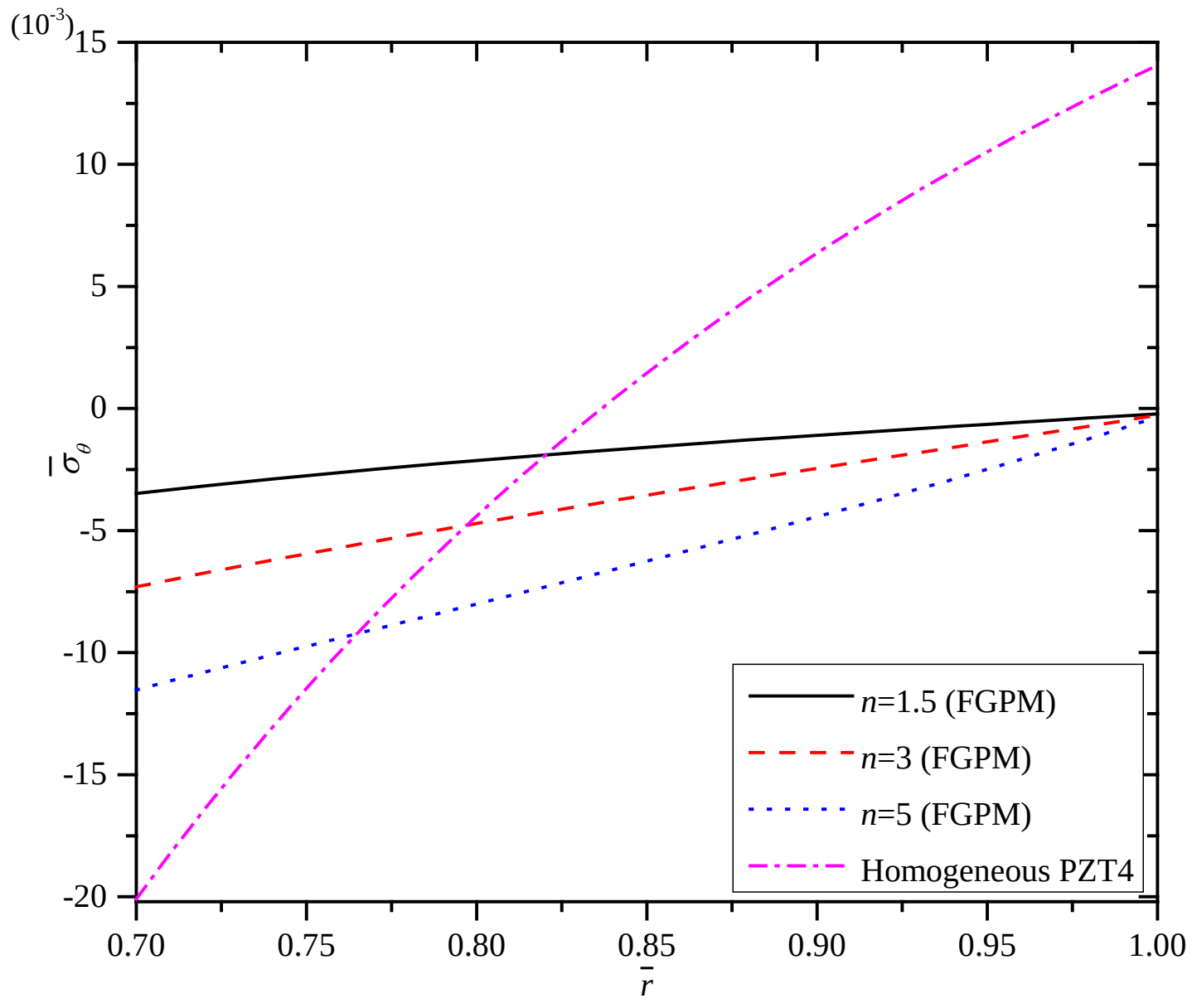

**Fig. 11.** Evolution of the circumferential stress with different parameter *n*

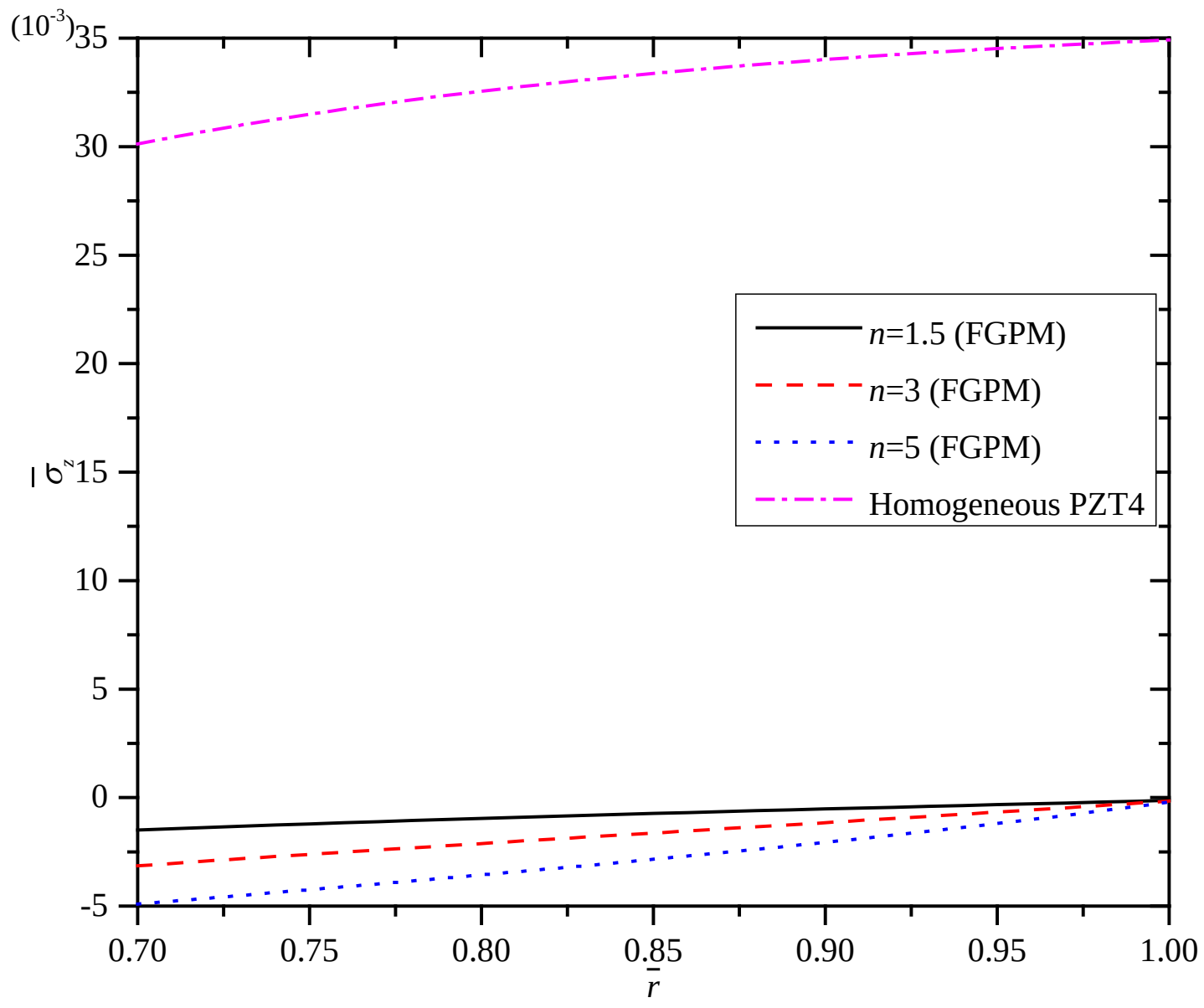

**Fig. 12.** Evolution of the axial stress with different parameter *n*

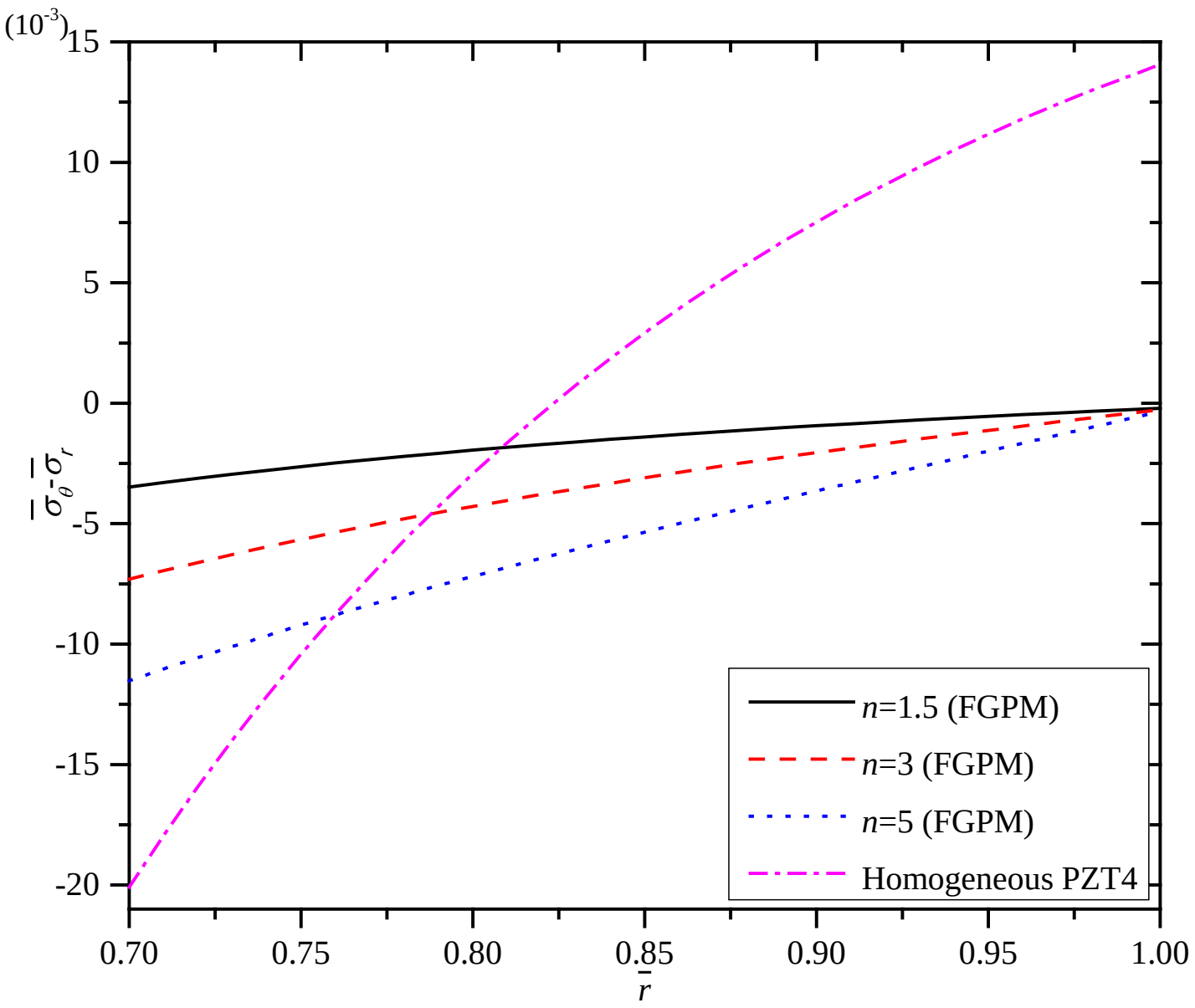

**Fig. 13.** Evolution of the difference between the circumferential stress and the radial stress with different parameter *n*

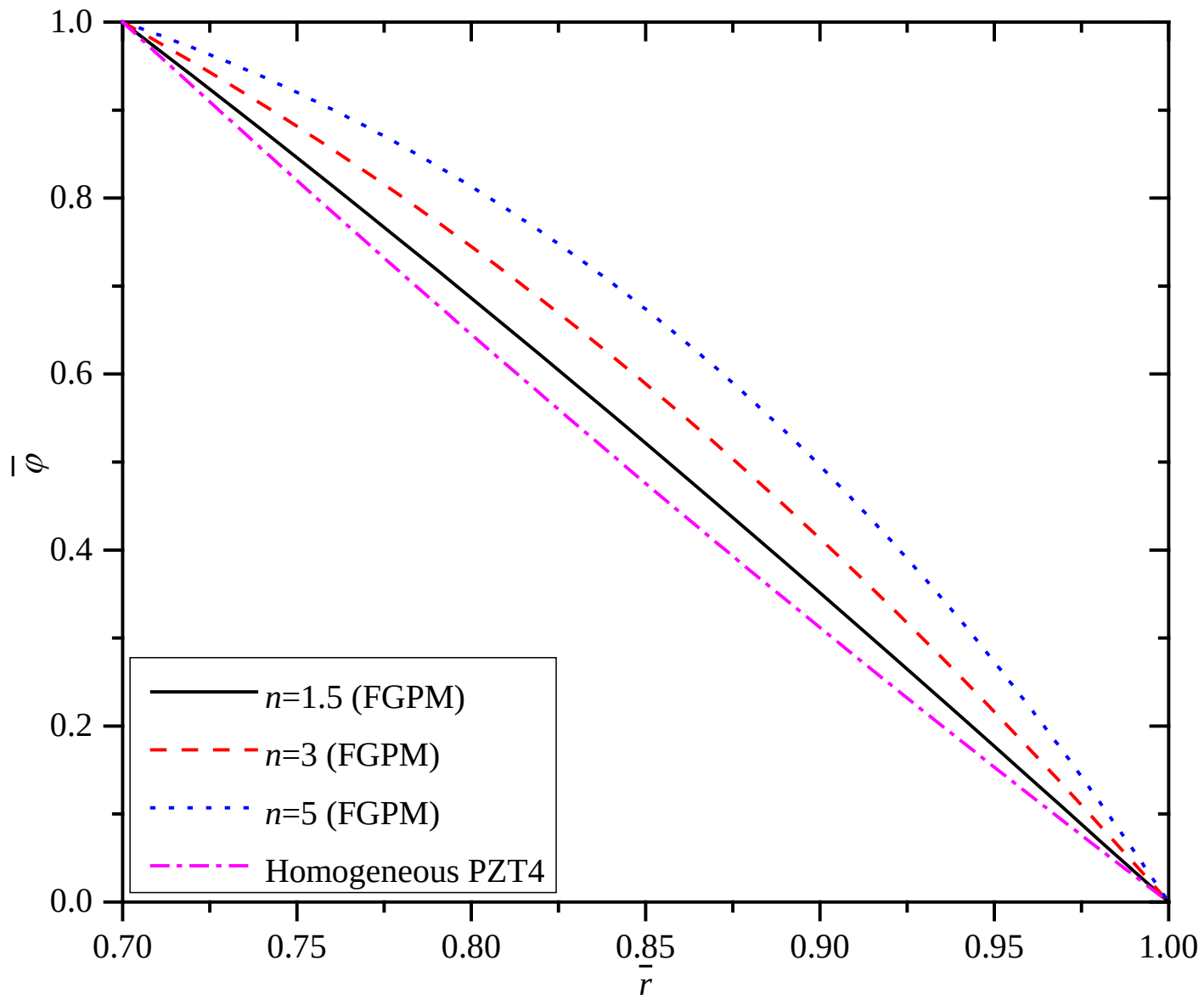

**Fig. 14.** Evolution of the electric potential with different parameter *n*